\documentclass[showpacs,preprintnumbers,amsmath,amssymb]{revtex4}
\usepackage{graphicx}% Include figure files
\usepackage{dcolumn}% Align table columns on decimal point
\usepackage{bm}% bold math
\usepackage{natbib}
\usepackage[dvips]{color}

\begin{document}

%\preprint{APS/123-QED}

\title{The Weibull - log Weibull transition of the Interoccurrence time statistics \\ in the two-dimensional Burridge-Knopoff Earthquake model}% Force line breaks with \\

\author{Tomohiro Hasumi}
 \email{t-hasumi.1981@toki.waseda.jp}
\author{Takuma Akimoto}
 \email{akimoto@aoni.waseda.jp}	
\author{Yoji Aizawa}
 \email{aizawa@waseda.jp}
\affiliation{Department of Applied Physics, Advanced School of Science and Engineering, Waseda University, Tokyo 169-8555, Japan}
%Lines break automatically or can be forced with \\

\date{\today}% It is always \today, today,
             %  but any date may be explicitly specified

\begin{abstract}
In analyzing synthetic earthquake catalogs created by a two-dimensional Burridge-Knopoff model, we have found that a probability distribution of the interoccurrence times, the time intervals between successive events, can be described clearly by the superposition of the Weibull distribution and the log-Weibull distribution. 
In addition, the interoccurrence time statistics depend on frictional properties and stiffness of a fault and exhibit the Weibull - log Weibull transition, which states that the distribution function changes from the log-Weibull regime to the Weibull regime when the threshold of magnitude is increased. 
%On the basis of renewal theoretical analysis, the occurrence of events in this model is completely stationary. 
%\textcolor{blue}{As the stiffness parameters are increased, the pure log-Weibull regime becomes small and then disappear. However, the dominant distribution of the interoccurrence time changes from the log-Weibull distribution to the Weibull distribution, which is in agreement with the observed features.}
%The features reproduce the interoccurrence time statistics in nature by analyzing the Japan earthquake data so that this model sufficiently reproduces the interoccurrence time statistics. 
We reinforce a new insight into this model; the model can be recognized as a mechanical model providing a framework of the Weibull - log Weibull transition.
\end{abstract}

% PACS, the Physics and Astronomy
                             % Classification Scheme.
%\keywords{Suggested keywords}%Use showkeys class option if keyword
                              %display desired
\maketitle

\section{\label{intro}Introduction}
Earthquakes are phenomena exhibiting great complexity and strong intermittency.  
Statistical mechanical approaches are applied to understand the complex fault systems~\cite{Rundle:RG2003}. 
Although the source mechanism of earthquakes is still open, statistical properties of earthquakes are well described by some empirical laws, such as, the Gutenberg-Richter (GR) law~\cite{Gutenberg:AGeo1956} and the Omori law for aftershocks~\cite{Omori:IUT1894}. 
The statistical properties of time intervals between successive earthquakes, hereinafter called the interoccurrence times, have been paid much attention and discussed~\cite{Corral:PRL2004, Lindman:PRL2005, Shcherbakov:PRL2005} since Bak {\it et al.} proposed the scaling law by analyzing Southern California earthquakes~\cite{Bak:PRL2002}. 
It is shown that the interoccurrence time distribution for the earthquakes occurring on a single fault can be described by the Weibull distribution, and that the Weibull exponent increases with the increase of the magnitude threshold~\cite{Abaimov:GJI2007}.\par

Very recently, in analyzing the Japan earthquake data, we have proposed a statistical feature of interoccurrence times, which states that the probability distribution can be definitely written by the superposition of the Weibull distribution and the log-Weibull distribution~\cite{Hasumi:condmat2008}. 
We have reinforced this view that the interoccurrence time statistics show the Weibull - log Weibull transition, which means that Weibull statistics and log-Weibull statistics coexist in the interoccurrence time statistics, and the dominant distribution function then changes from the log-Weibull distribution to the Weibull distribution as the threshold of magnitude is increased. 
It was found that the crossover magnitude from the superposition domain to the Weibull domain depends on the location on which we focused.   
These features are also observed by analyzing the Southern California and Taiwan earthquake data~\cite{Hasumi:inprep2008}. 
Meanwhile, the Weibull - log Weibull transition also appeared in dynamical systems~\cite{Akimoto:PTP2005}. 
However, the theoretical background and interpretation of this transition remain to be developed.\par

Many earthquake models have been proposed and simulated to infer the source mechanism of earthquakes and to discuss whether the model reproduces the statistical properties of earthquakes. 
The Burridge-Knopoff (BK) model~\cite{Burridge:BSSA1967} is one of the theoretical models of earthquakes. 
Recent works on this model have been studied by many authors~\cite{Mori:PRE2008, Xia:PRE2008, Lippiello:EPL2005, Abaimov:NPG2007, Hasumi:PRE2007}. 
Abaimov {\it et al.} showed that interoccurrence time distributions depended on the stiffness of the system by simulating the one-dimensional (1D) BK model with the static-dynamic friction law~\cite{Abaimov:NPG2007}.  
They found that for a small stiffness the interoccurrence time distribution exhibits an exponential distribution, while for a large stiffness the interoccurrence time distribution restricted to system-wide events obeys the Weibull distribution. 
The system-wide events were defined as the events where all blocks slip during an event.
In the 2D BK model, the present author demonstrated that the cumulative distribution of the interoccurrence time can be described by the power law~\cite{Hasumi:PRE2007}, which reproduces the observed behavior~\cite{Abe:PA2005}. 
To the best of our knowledge, a mechanical model which exhibits the Weibull - log Weibull transition has not been reported. \par

In this study, we attempt to understand how the Weibull - log Weibull transition and the crossover magnitude are influenced by a change in the major physical quantities, such as frictional properties and stiffness of a fault. 
In addition, it is worth discussing whether the BK model shows the Weibull - log Weibull transition. 
Thus we study the interoccurrence time statistics produced by the 2D BK model by changing the friction and stiffness parameters and the threshold of magnitude. 
As a result, the interoccurrence time statistics exhibit the Weibull - log Weibull transition. 
%, which is compatible with the observed features of real earthquakes~\cite{Hasumi:condmat2008, Hasumi:inprep2008}.   
%Furthermore, the sequence of earthquakes in this model is completely stationary on the basis of the theoretical renewal analysis. 

\section{\label{model}The Model and the method}
\begin{figure*}[]
\begin{center}
\includegraphics[width=.4\linewidth]{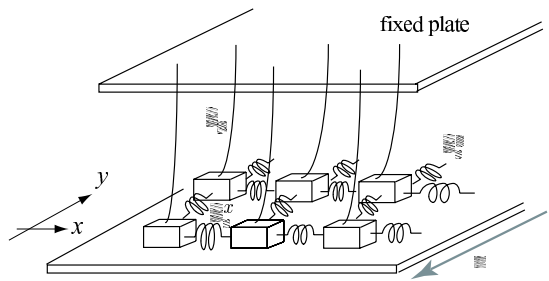}
\end{center}
\caption{Schematic illustration of the 2D Burridge-Knopoff (BK) model.}
\label{BK_model}
\end{figure*}
We have simulated the 2D BK model describing the relative motion of faults. 
As shown in fig.~\ref{BK_model}, this model consists of two plates, three kinds of springs corresponding to the stress acting on the surface of faults, and blocks whose mass is $m$ interconnected by the linear spring of elastic constants $k_c^x $ and $k_c^y$. 
The blocks are connected with a fixed plate by the spring with spring constant $k_p$. 
We assume that the slipping direction is restricted to the $y$-direction. 
The equation of motion at site $(i,j)$ can be expressed by
\begin{eqnarray}
\ddot{y}_{i,j} = k_c^x(y_{i+1,j}+y_{i-1,j}-2y_{i,j})  +  k_c^y(y_{i,j-1}+y_{i,j+1}-2y_{i,j})  - k_p y_{i,j} -  F\left(\left[v+\dot{y}_{i,j}\right]\right),
\label{eqn:eqm}
\end{eqnarray}
where $y_{i,j}$ is the displacement, $v$ is the plate velocity, and $F$ is the dynamical friction force as a function of the velocity of the block and $v$. 
In order to rewrite this equation in a dimensionless form, we define the dimensionless position $U$, the friction function $\phi$, and time $t'$ as 
\begin{eqnarray}
U_{i,j}= y_{i,j}/D_0 = y_{i,j}/(F_0/k_p),  F(\dot{y}_{i,j}) = F_0\phi(\dot{y}_{i,j}/v_1),  t' = \omega_pt = \sqrt{k_p/m}\; t, \nonumber
\end{eqnarray}
where $F_0$ is the maximum friction force and $v_1$ is a characteristic velocity. 
Substituting these parameters for $y, t,$ and $F$, we obtain the non-dimensional equation of motion,
\begin{eqnarray}
\ddot{U}_{i,j} = l_x^2(U_{i+1,j}+U_{i-1,j}-2U_{i,j})  +  l_y^2(U_{i,j-1}+U_{i,j+1}-2U_{i,j})  - U_{i,j} -  \phi \left(2\gamma \left[\nu+\dot{U}_{i,j}\right]\right),
\label{eqn:noneqm}
\end{eqnarray}
where dots indicate derivatives with respect to $t'$.  
$l_x (= \sqrt{k_c^x/k_p})$ and $l_y (= \sqrt{k_c^y/k_p})$ are the stiffness in the $x$ and $y$ directions, respectively. 
$\nu$ represents the dimensionless loading velocity, which stands for the ratio of the plate velocity to the maximum slipping velocity $\omega_p D_0$. 
$\phi$ is the dynamical friction function. As the form of $\phi$, we use a velocity-weakening friction force introduced in Ref.~\cite{Carlson:PRA1991}, 
\begin{eqnarray}
\phi (\dot{U}) = \left\{
\begin{array}{ll}
(-\infty, 1] & \dot{U}=0,\\
{\displaystyle \frac{(1-\sigma)}{\{1+2\gamma[\dot{U}/(1-\sigma)]\}}} & \dot{U}>0,
\end{array}
\right.
\label{friction_function}
\end{eqnarray}
where $\gamma$ is a decrement in the dynamical friction force, and $\sigma$ is the difference between the maximum friction force and the dynamical friction force $\phi(0)$. 
To forbid a back slip, which means that a block slips in the $-y$ direction, $\phi (\dot{U})$ ranges from $-\infty$ to 1, arbitrarily. 
This system is governed by five parameters, $l_x, l_y, \sigma, \nu$, and $\gamma$. Throughout this work $\sigma$ and $\nu$ are set at 0.01, because our previous work reported that the optimal parameters of this model are estimated to be $l_x^2=1, l_y^2=3, \sigma = 0.01, \nu=0.01$, and $\gamma=3.5$ in view of the reproduction of statistical properties of earthquakes, for instance, the GR law with $b=1$, the Zipf-Mandelbrot type power law for interoccurrence time statistics~\cite{Hasumi:PRE2007}.
%Therefore the results in this paper are for the optimal case with the system size $(N_x, N_y)=(100,25)$. 
We calculate Eqs.~(\ref{eqn:noneqm}) and (\ref{friction_function}) under the free boundary condition with the 4th order Runge-Kutta algorithm with the system size $(N_x, N_y)=(100,25)$.

In this model, a time when a block slips for the first time during an event is considered as the earthquake occurrence time. 
The $n$th interoccurrence time is defined as $\tau_n = t_{n+1}-t_{n}$, where $t_{n}$ and $t_{n+1}$ are the occurrence times of the $n$th and $n+1$th earthquake, respectively. 
The interoccurrence statistics are then studied by changing $l_x^2, l_y^2, \gamma$, and the threshold magnitude $m_c$. 
A seismic magnitude $m$ in this model is defined as $m = \log_{10} \left( \sum _{i,j}^n  \delta u_{i,j} \right)/1.5$. 
$\delta u_{i,j}$ stands for the total displacement at site $(i,j)$ during an event and $n$ is the number of slipping blocks. \par
%\section{Applicable distribution of inter-occurrence times}
One of our goals in this work is to detect the distribution density function of the interoccurrence time $P(\tau)$. 
For this purpose, we selected several kinds of distribution for $P(\tau)$; the Weibull distribution $P_{w}$~\cite{Abaimov:PEP2008, Hasumi:condmat2008}, the log-Weibull distribution $P_{lw}$~\cite{Hasumi:condmat2008, Huillet:EPJB1999}, the power law $P_{pow}$~\cite{Abe:PA2005}, the gamma distribution $P_{gam}$ (in the case of $\delta=1$ in the paper \cite{Corral:PRL2004}), and the log normal distribution $P_{ln}$~\cite{Abaimov:PEP2008, Matthews:BSSA2002} which are defined by,
\begin{eqnarray}
P_w(\tau) = {\displaystyle \left(\frac{\tau}{\beta_1} \right)^{\alpha_1 -1} \frac{\alpha_1}{\beta_1} \exp \left[-\left(\frac{\tau}{\beta_1}\right)^{\alpha_1} \right]}, \nonumber\\
P_{lw}(\tau) = {\displaystyle \frac{(\log (\tau /h))^{\alpha_2 -1}}{(\log \beta_2 )^{\alpha_2}} \frac{\alpha_2}{\tau} 
\exp \left[-\left(\frac{\log (\tau /h)}{\log \beta_2}\right)^{\alpha_2}  \right]}, \nonumber\\
P_{pow}(\tau) = {\displaystyle \frac{\beta_3(\alpha_3 -1)}{(1+\beta_3 \tau)^{\alpha_3}}}, \nonumber\\
P_{gam}(\tau) = {\displaystyle \tau^{\alpha_4 -1} \frac{\exp{(-\tau/\beta_4)}}{\Gamma(\alpha_4) {\beta_4}^{\alpha_4}}},\nonumber\\
P_{ln}(\tau) = {\displaystyle \frac{1}{\tau \beta_5\sqrt{2\pi}}\exp \left[-\frac{(\ln(\tau)-\alpha_5 )^2}{2\beta_5 ^2}\right]}, \nonumber
\end{eqnarray}
where $\alpha_{i}, \beta_{i}$ and $h$ are constants and characterize the distribution. 
In this time, $h$ is fixed at 0.5. 
$\Gamma(x)$ is the gamma function. 
$i$ stands for an index number; $i=1, 2, 3, 4$, and 5 correspond to the Weibull distribution, the log-Weibull distribution , the power law, the gamma distribution , and the log normal distribution, respectively. 
Note that these distributions have been used as a fitting function of $P(\tau)$. 
We will comment on the log-Weibull distribution. 
This distribution is constructed by the logarithmic modification of the Weibull distribution. 
Generally speaking, the tail of the log-Weibull distribution is much longer than that of the Weibull distribution. 
Previously, the log-Weibull distribution was derived from the chain-reaction model introduced by Huillet and Raynaud, and they then applied the log-Weibull distribution to fit the magnitude data in France and Japan~\cite{Huillet:EPJB1999}. 
To maintain the statistical accuracy, we analyze the interoccurrence times using at least 100 events. 
In this work, the root mean square (rms) test and the Kolmogorov-Smirnov test are used in order to determine the most suitable distribution function. 
The rms value is defined as 
\begin{eqnarray}
rms = \sqrt{\frac{\sum_{i=1} ^{n'} (x_i - x_i')^2}{n'-k}},
\end{eqnarray}
where $x_i$ and $x_i'$ are actual data and predicted data derived from the best-fit curve, respectively. 
$n'$ is the number of data plots and $k$ is the number of fitting parameters. 
It is well-known that the preferred distribution yields the smallest rms value. 
We calculate the rms value by use of the cumulative distribution function obtained from the numerical data to reduce the statistical fluctuations. 
Then the Kolmogorov-Smirnov test is performed in order to provide a more accurate confidence level, where the maximum deviation statistic $D_{KS}$ which is defined by
\begin{eqnarray}
D_{KS} = \max_i |y_i -y_i'|,
\end{eqnarray}
where $y_i$ and $y_i'$ stand for the actual data of the cumulative distribution and the data estimated from the fitting distribution, respectively. 
It is well recognized that the preferred distribution has the smallest value of $D_{KS}$.
%\textcolor{blue}{Note that residuals are weighted equally when we perform the data-fitting.} 

\section{Results and Discussion}

\begin{figure*}[]
\begin{center}
\includegraphics[width=.95\linewidth]{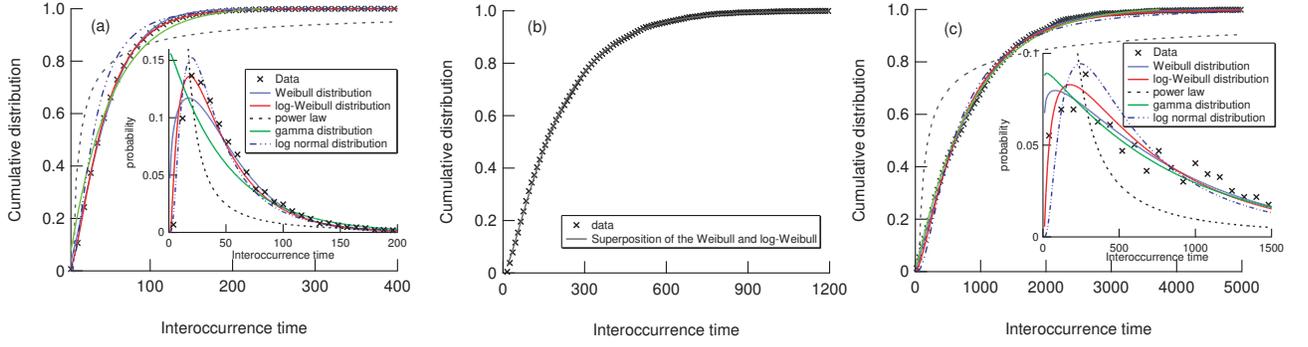}
\end{center}
\caption{The cumulative distribution of interoccurrence times in the 2D BK model for different $m_c$. (a), (b), and (c) correspond respectively to the log-Weibull regime $m_c=0.3$, superposition regime $m_c=0.8$, and the Weibull regime $m_c=1.1$. Insets represent the probability density function. In (b), the solid line stands for the optimal parameter fitting by eq.~(\ref{conjecture}), where $\alpha_1=1.19 \pm 0.005, \beta_1=2.21\times 10^{2} \pm 0.56, \alpha_2 = 7.19 \pm 0.03, \beta_2 = 4.31\times 10^{2} \pm 0.89$, and $p=0.40 \pm 0.04$, and the symbols $(\times)$ are the numerical data obtained from 4609 events and 147 data points.}
\label{cumu}
\end{figure*}

\begin{table}
\caption{\label{table1}The results of the rms value, $D_{KS}$, and fitting parameters for different distribution functions.}
\begin{center}
\begin{tabular}{c|c|c|c|c|c|c|c}
 &distribution&$\alpha_i$ &$\beta_i$&rms & $\ln$ rms & $D_{KS}$ & ln $D_{KS}$ \\
\hline
\hline
  & $P_{w}~(i=1)$ & 1.08$\pm$ 0.007 & 8.29$\times 10^{2}\pm$ 3.34 & 6.4$\times 10^{-3}$ & $-5.05$ & 0.019 & $-3.96$ \\
 $m_c=1.1$ & $P_{lw}~(i=2)$ & 8.05$\pm$ 0.10 & 1.62$\times 10^{3} \pm$ 13.3 & 1.3$\times 10^{-2}$ & $-4.34$ & 0.040 & $-3.21$ \\
 1138 events & $P_{pow}~(i=3)$ & 1.56 $\pm$ 0.04 & 6.36$\times 10^{1} \pm$ 8.42 & 1.3$\times 10^{-1}$ & $-2.07$ & 0.35 & $-1.05$ \\
 70 data points & $P_{gam}~(i=4)$ & 1.02$\pm$ 0.002 & 8.21$\times 10^{2} \pm$ 4.91 & 9.5$\times 10^{-3}$ & $-4.65$ & 0.026 & $-3.65$ \\
  & $P_{ln}~(i=5)$ & 6.33$\pm$ 0.01 & 0.92$\pm$ 0.04 & 2.4$\times 10^{-2}$ &  $-3.71$ & 0.077 &  $-2.56$ \\
\hline
  & $P_{w}~(i=1)$ & 1.31$\pm$ 0.01 & 5.02$\times 10^{1}\pm$ 0.27 & 7.4$\times 10^{-3}$ & $-4.90$ & 0.036 & $-3.32$ \\
 $m_c=0.3$ & $P_{lw}~(i=2)$ & 5.92$\pm$ 0.007 & 9.74$\times 10^{1} \pm$ 0.85 & 8.5$\times 10^{-4}$ & $-7.08$ & 0.0027 & $-5.91$ \\
 19545 events & $P_{pow}~(i=3)$ & 1.70 $\pm$ 0.04 & 5.57$\pm$ 0.55 & 9.5$\times 10^{-2}$  & $-2.35$ & 0.035 & $-1.05$ \\
 80 data points & $P_{gam}~(i=4)$ & 1.01$\pm$ 0.005 & 4.86$\times 10^{1} \pm$ 0.90 & 2.3$\times 10^{-2}$ & $-3.78$ & 0.11 & $-2.23$ \\
  & $P_{ln}~(i=5)$ & 3.59$\pm$ 0.005 & 0.79$\pm$ 0.01 & 6.5$\times 10^{-3}$ & $-5.03$ & 0.025 & $-3.69$ \\
\hline
\end{tabular}
\end{center}
\end{table}

We attempt to trace a change in the interoccurrence time statistics by changing $m_c$. 
Our previous work~\cite{Hasumi:PRE2007} was mainly focused on the no-threshold case, $m_c=-\infty$. 
The cumulative distributions of the interoccurrence times for different $m_c$ are displayed in Fig.~\ref{cumu} for $l_x^2=1, l_y^2=3, \sigma = 0.01, \nu=0.01$, and $\gamma=3.5$.
For (a), (b), and (c), $m_c$ is set at $m_c=0.3$, 0.8, and 1.1, respectively. 
The results of the rms value and $D_{KS}$ for a different distribution function for $m_c=1.1$ and $m_c=0.3$ are listed in Table~\ref{table1}.  
As can be seen from the table, for small $m_c$, (e.g., $m_c=0.3$) the interoccurrence time distribution obeys the log-Weibull distribution, whereas for a large $m_c$ (e.g., $m_c=1.1$) the Weibull distribution is preferred. 
However, the fitting accuracy of the Weibull distribution becomes worse when $m_c$ is decreased. 
At the same time, the fitting accuracy of the log-Weibull distribution becomes worse as $m_c$ is increased.
Hence, we think that for the intermediate case (b), the distribution can be described by the superposition of the Weibull distribution and the log-Weibull distribution because the Weibull components and the log-Weibull components of  $P(\tau)$ do not disappear suddenly. 
Actually, $P(\tau)$ is well fitted by the superposition of the Weibull distribution and the log-Weibull distribution.  
Taken together, the probability distribution of the interoccurrence time can be expressed explicitly by the following;
\begin{eqnarray}
P(\tau)= p\times {\text {Weibull distribution}} + (1-p)\times {\text{log-Weibull distribution}},
\label{conjecture}
\end{eqnarray}
where $p$ means the rate of the Weibull distribution in the range, $0 \le p \le 1$ and depends on $m_c$. 
As for $p=1$, $P(\tau)$ obeys the Weibull distribution,  whereas for $p = 0$, $P(\tau)$ follows the log-Weibull distribution.  
$P(\tau)$ is characterized by five parameters, $\alpha_1, \alpha_2, \beta_1, \beta_2$, and $p$, while for the pure Weibull and log-Weibull regimes, the number of fitting parameters is two; $\alpha_1$ and  $\beta_1$ for $P_{w}$, and $\alpha_2$ and $\beta_2$ for $P_{lw}$. 
It is reported that the interoccurrence time distribution can be described by the superposition of the log-Weibull distribution and the Weibull distribution in the same manner of eq.~(\ref{conjecture}) by analyzing the Japan, California, and Taiwan earthquake data~\cite{Hasumi:condmat2008, Hasumi:inprep2008}. \par 

\subsection{Friction parameter $\gamma$ dependence} 

\begin{figure*}[]
\begin{center}
\includegraphics[width=.95\linewidth]{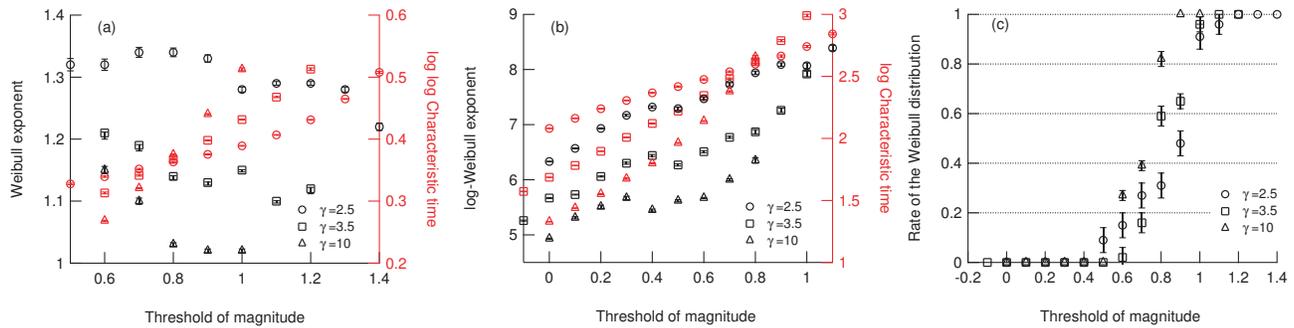}
\end{center}
\caption{Change in fitting parameters as a function of $m_c$ for different $\gamma$ values while fixing $l_x^2=1$ and $l_y^2=3$ $(\circ: \gamma=2.5, ~\square: \gamma=3.5,$ and $\triangle: \gamma=10)$. The Weibull components, the log-Weibull components, and the rate of Weibull distribution are shown in (a), (b), and (c), respectively.}
\label{friction}
\end{figure*}
Next, we focus on a change in fitting parameters by varying the friction parameter $\gamma$ and $m_c$. 
For this purpose, we allow $\gamma$ to range from 1.0 to 10 while fixing $l_x^2=1$ and $l_y^2=3$.  
Figure.~\ref{friction} shows the dependence of the fitting parameters on $m_c$ for different $\gamma$ values ($\gamma=2.5~(\circ), 3.5~(\square)$, and 10~$(\triangle)$).  
As can be seen from Fig.~\ref{friction} (a), the Weibull exponent $\alpha_1$ gradually decreases, and the characteristic  time $\beta_1$ increases double exponentially. 
Note that $\beta_1$ increases more rapidly for a large value of $\gamma$. 
As shown in Fig.~\ref{friction} (b), the log-Weibull exponent $\alpha_2$ increases linearly, and the characteristic time $\beta_2$ increases exponentially with $m_c$. 
The rate of the Weibull distribution $p$ ranges from 0 to 1 as $m_c$ is increased, indicating the fact that the distribution $P(\tau)$ exhibits the Weibull - log Weibull transition (see Fig.~\ref{friction} (c)). 
The transition magnitude from the log-Weibull regime to the superposition regime and from the superposition regime to the Weibull regime is denoted respectively by $m_c^{*}$ and $m_c^{**}$ depending on $\gamma$; $m_c^{*}=0.4$ and $m_c^{**}=1.2$ for $\gamma=2.5$, $m_c^{*}=0.5$ and $m_c^{**}=1.1$ for $\gamma=3.5$, and $m_c^{*}=0.5$ and $m_c^{**}=0.9$ for $\gamma=10$. 
We conclude that the Weibull - log Weibull transition holds while changing $\gamma$.

\subsection{Stiffness parameters $l_x^2$ and $l_y^2$ dependence}
\begin{figure*}[]
\begin{center}
\includegraphics[width=.95\linewidth]{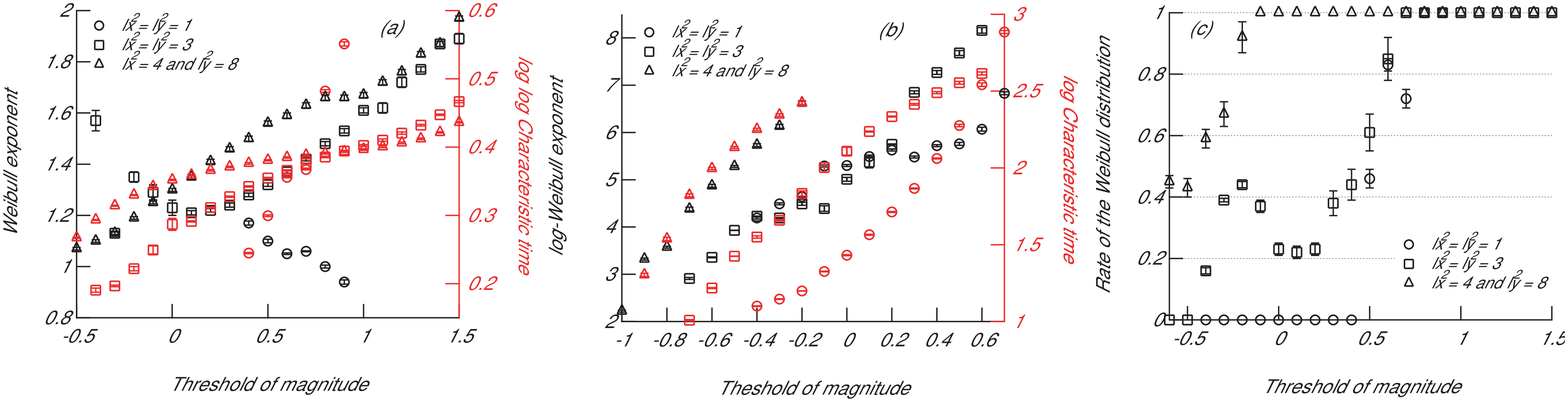}
\end{center}
\caption{The relation between fitting parameters and $m_c$ for different $l_x^2$ and $l_y^2$, whereas $\gamma=3.5$ $(\circ: l_x^2=l_y^2=1, \square: l_x^2=l_y^2=3$, and $\triangle: l_x^2=4$ and $l_y^2=8)$. The components of the Weibull distribution and the log-Weibull distribution are displayed in (a) and (b), respectively. In (c) the ratio of Weibull distribution is shown.}
\label{stiffness}
\end{figure*}
In the second performance of our simulation, we focus on the relation between interoccurrence times and stiffness parameters, $l_x^2$ and $l_y^2$. 
In order to achieve this, $\gamma$ is fixed at 3.5, while $l_x^2$ and $l_y^2$ are varied. 
As demonstrated in Fig.~\ref{stiffness} (a), the change of the Weibull exponent $\alpha_1$ can be classified into three types; first in the case of $l_x^2=l_y^2=1~(\circ)$, $\alpha_1$ gradually decreases as the $m_c$ is increased. 
Second, for $l_x^2=l_y^2=3~(\square)$, although $\alpha_1$ decreases with $m_c$ in the region $m_c \lessapprox 0.1$, $\alpha_1$ increases in the region $m_c \gtrapprox 0.1$. 
Finally, for $l_x^2=4$ and $l_y^2=8$ $(\triangle)$, $\alpha_1$ increases with the increase of $m_c$. 
Similar tendency is observed by use of the Southern California earthquake data~\cite{Abaimov:GJI2007}. 
$\beta_1$ expands double exponentially as $m_c$ increases. 
Figure~\ref{stiffness} (b) shows that the log-Weibull components, $\alpha_2$ and $\beta_2$ increase linearly and exponentially, respectively with $m_c$. 
Finally, as clearly seen from Fig.~\ref{stiffness} (c), the Weibull - log Weibull transition is observed because $p$ changes from 0 to 1 when $m_c$ is increased in the case, both stiffness parameters small enough, $l_x^2 \lessapprox 3$ and $l_y^2 \lessapprox 3$. 
The values of the transition magnitude, $m_c^{*}$ and $m_c^{**}$ depend on $l_x^2$ and $l_y^2$; $m_c^{*}=0.4$ and $m_c^{**}=0.8$ for $l_x^2=l_y^2=1$, and $m_c^{*}=-0.5$ and $m_c^{**}=0.6$ for $l_x^2=l_y^2=3$. 
It should be noted that for $l_x^2=4$ and $l_y^2=8$, the pure log-Weibull regime does not appear, because $p>0$ for any $m_c$. 
In this case, the interoccurrence time statistics contain both the Weibull and the log-Weibull components, and then the dominant distribution changes from the log-Weibull distribution to the Weibull distribution when the threshold $m_c$ is increased. 
The first transition point $m_c^{*}$ cannot be determined clearly, but the second transition point $m_c^{**}$ is estimated to be $m_c^{**}=-0.1$ (see Fig.~\ref{stiffness} (c)). \par

\subsection{System size dependence}
\begin{figure*}[]
\begin{center}
\includegraphics[width=.4\linewidth]{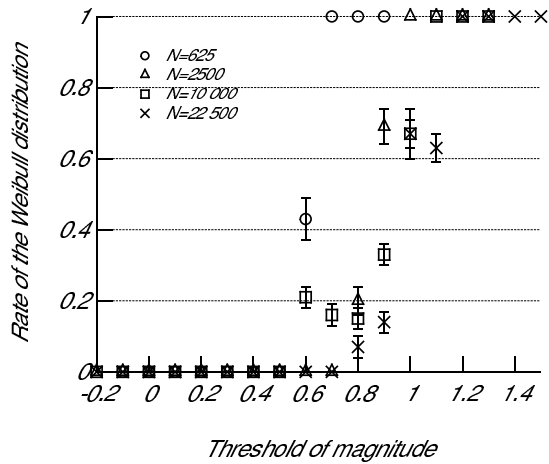}
\end{center}
\caption{The ratio of Weibull distribution $p$ as a function of $m_c$ for different numbers of blocks $N$ ($\circ:625$, $\triangle:2500$, $\square:10000$, and $\times:22500$).}
\label{size}
\end{figure*}

The system size dependence of the interoccurrence times is studied to discuss the finite size effect. 
In this time, the number of blocks $N$ is changed from $625~(25\times 25)$ to $22500~(150\times 150)$, while other parameters are fixed at $l_x^2=1, l_y^2=3$, and $\gamma=3.5$; $N$ is taken to be $N=625~(25\times 25)$, $N=2500~(50\times 50)$, $N=10000~(100\times 100)$, and $N=22500~(150\times 150)$. 
As clearly seen from Fig.~\ref{size}, $p$ changes from 0 to 1 as $m_c$ is increased, suggesting the fact that the Weibull - log Weibull transition appears in all cases. 
Transition magnitudes, $m_c^{*}$ and $m_c^{**}$ are then evaluated; $m_c^{*}=0.5$ and $m_c^{**}=0.7$ for $N=625~(\circ)$, $m_c^{*}=0.7$ and $m_c^{**}=1.0$ for $N=2500~(\triangle)$, $m_c^{*}=0.5$ and $m_c^{**}=1.1$ for $N=10000~(\square)$, and $m_c^{*}=0.7$ and $m_c^{**}=1.2$ for $N=22500~(\times)$. 
%In addition, the the transition magnitude from the superposition regime to the Weibull one depends on the system size, which can be  reproduced the observed behavior of real earthquakes~\cite{} 
Thus, we conclude that the interoccurrence time statistics, especially the Weibull - log Weibull transition, are retained for a large system size.

\subsection{Origin of the log-Weibull distribution} 
\begin{figure*}[]
\begin{center}
\includegraphics[width=.4\linewidth]{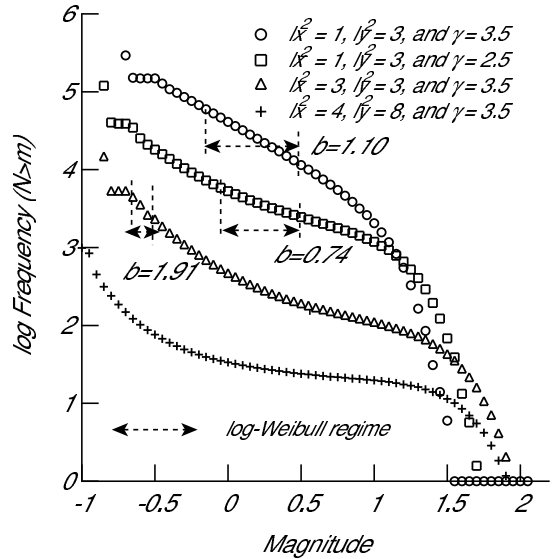}
\end{center}
\caption{The cumulative number of earthquakes as a function of magnitude obtained from the 2D BK model. Circles ($\circ$), squares ($\square$), triangles ($\triangle$), and plus signs ($+$) correspond to the case of  $l_x^2=1$, $l_y^2=3$, and $\gamma=3.5$, $l_x^2=1, l_y^2=3,$ and $\gamma=2.5$, $l_x^2=3, l_y^2=3$, and $\gamma=3.5$,  and  $l_x^2=4, l_y^2=8$, and $\gamma=3.5$. The log-Weibull region, where $p=0$ is denoted by an arrow.  All the plots except for the case of $l_x^2=1, l_y^2=3$, and $\gamma=3.5$ are shifted vertically for clarity.}
\label{magnitude}
\end{figure*}

As we mentioned, the probability distribution of interoccurrence times can be described evidently by the eq. (\ref{conjecture}). 
However, as $l_x^2$ and $l_y^2$ are increased, the pure log-Weibull regime becomes small (e.g., $l_x^2=l_y^2=3$) and then disappears (e.g., $l_x^2=4$ and $l_y^2=8$). 
In this study we deduce the role of the log-Weibull distribution in view of the magnitude distribution. 
The cumulative number of earthquakes $N$ whose magnitude is greater than or equal to $m$ $(N \ge m)$ as a function of $m$ for different parameters produced by the 2D BK model are presented in Fig.~\ref{magnitude}.
The arrow in Fig.~\ref{magnitude} stands for the pure log-Weibull regime where $p=0$. 
As shown this figure, when the distribution obeys the power law, the pure log-Weibull regime can be observed, suggesting the conjecture that the origin of the log-Weibull distribution in the 2D BK model is related to the power law magnitude distribution. 
Note that the parameter region, where the magnitude distribution obeys the power law globally, is limited. 
For $l_x^2=1, l_y^2=3$, and $\gamma=3.5$, the power law exponent $b$ is $b=1.10$, which is the similar to that value obtained from the earthquake data, $b \sim 1$.  

\subsection{The onset mechanism of the Weibull distribution}
\begin{figure*}[]
\begin{center}
\includegraphics[width=.4\linewidth]{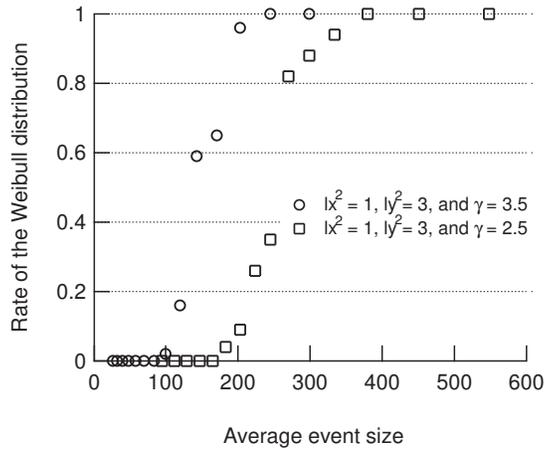}
\end{center}
\caption{The relation between the average size event and the ratio of the Weibull distribution. Circles ($\circ$) and squares ($\square$) correspond to the case of  $l_x^2=1$, $l_y^2=3$, and $\gamma=3.5$ and $l_x^2=1, l_y^2=3,$ and $\gamma=2.5$, respectively.}
\label{size_p}
\end{figure*}

Next, we focus on the onset mechanism of the Weibull distribution from the viewpoint of the average event size. 
Here, the size of an event is defined as the number of slipping blocks during the event. 
It was shown that the time-interval distribution of the system-wide events obeys the Weibull distribution~\cite{Abaimov:NPG2007, Abaimov:PEP2008}.
This enables us to the conjecture that the Weibull distribution is induced from the enhancement of the average event size, $\bar{S}$. 
This conjecture is supported in Fig.~\ref{size_p}, where we show the relation between the ratio $p$ and the average event size $\bar{S}$, and the parameter values correspond to the cases treated in Fig.~\ref{friction} (c). % in the case of $l_x^2=1$, $l_y^2=3$, and $\gamma=3.5$ and $l_x^2=1, l_y^2=3,$ and $\gamma=2.5$ in Fig.~\ref{size_p}.
%As a result, $p$ gradually increases as $\bar{S}$ increases indicating that this result supports the hypothesis.

%\subsection{Stationarity of events}
%Finally, the stationarity of earthquakes is discussed. 
%We examine thjs by means of the probability distribution of the inter-occurrence time based on the renewal theory. 
%The method is the same as that in our previous study~\cite{Hasumi:condmat2008}. 
%As a result, the sequence of earthquakes in this model is completely stationary because the first and the second moments of the Weibull distribution and the log Weibull distribution derived from our results are finite. 
%This result is the same as the observed behavior of real earthquakes~\cite{Hasumi:condmat2008, Hasumi:inprep2008}. 
%It should be noted that for $\alpha_2 <1$ these moments of the log Weibull distribution are infinite. \par

\section{Concluding Remarks}
We analyzed the interoccurrence time statistics produced by the 2D BK model by varying the dynamical parameters, $l_x^2, l_y^2$, and $\gamma$, for different thresholds of magnitude $m_c$. 
It is found that the probability distribution of the interoccurrence time can be described by the superposition of the Weibull distribution and the log-Weibull distribution. 
The statistics depend on $l_x^2, l_y^2$, and $\gamma$ and exhibit the Weibull - log Weibull transition, which states that the distribution function changes from the log-Weibull regime to the Weibull regime when $m_c$ is gradually increased. 
As $l_x^2$ and $l_y^2$ are increased, the log-Weibull domain becomes small and then disappears. 
On the contrary, the interoccurrence time distribution of large magnitude events always shows the Weibull distribution. 
%It is demonstrated that the occurrence of earthquakes is completely stationary by examining the distribution function on the basis of renewal theoretical analysis. 
%Our results presented here reproduce the observed behavior of the interoccurrence time statistics~\cite{Abaimov:GJI2007, Hasumi:condmat2008, Hasumi:inprep2008} by analyzing the earthquake data of Japan, Southern California, and Taiwan.  
%Thus it is possible to propose statistical properties of earthquakes in the stationary regime by analogy with the simulation studies of this model. 
Additionally, we proposed a new insight into the 2D BK model; the model can be recognized as a mechanical model exhibiting the Weibull - log Weibull transition.  
In this study, it is shown for the first time that the interoccurrence time distribution exhibits the log-Weibull distribution, reinforcing the view that the long-range correlation hides in the 2D BK model. 
Thus, we will focus on the analysis of the spatio-temporal correlation in future. %need to analyze the interoccurrence time statistics by considering the spatio-temporal correlation. 
In the BK model, fault dynamics are modeled as the stick-slip motion so that we infer that there is a possibility that other physical systems exhibiting the stick-slip motion might show the Weibull - log Weibull transition.   
We believe that this study provides a clue to the origin and the interpretation of this transition.

%The author would like to thank Professor Y. Aizawa and Dr. T. Akimoto for discussions. 
\begin{acknowledgments}
This work is partly supported by the Sasagawa Scientific Research Grant from The Japan Science Society. 
TH is grateful for research support from the Japan Society for the Promotion of Science (JSPS) and the Earthquake Research Institute cooperative research program at the University of Tokyo. 
Thanks are also extended to Dr. Sergey Abaimov and the three anonymous reviewers for improving the manuscript. 
\end{acknowledgments}

\end{document}